Article type: Full paper

# Title: Single-crystalline aluminum film for ultraviolet plasmonic nanolasers

*Bo-Tsun Chou, Yu-Hsun Chou, Yen-Mo Wu, Yi-Chen Chung, Wei-Jen Hsueh, Shih-Wei Lin, Tien-Chang Lu, Tzy-Rong Lin, and Sheng-Di Lin\**

((Optional Dedication))
Bo-Tsun Chou, Shih-Wei Lin, and Prof. Sheng-Di Lin*
Department of Electronics Engineering and Institute of Electronics,
National Chiao Tung University, Hsinchu, Taiwan
Address: 1001 University Road, Hsinchu, 30010, Taiwan
E-mail: sdlin@mail.nctu.edu.tw

Yu-Hsun Chou
Institute of Lighting and Energy Photonics, National Chiao Tung University, Tainan, Taiwan

Yen-Mo Wu and Prof. Tien-Chang Lu
Department of Photonics, National Chiao Tung University, Hsinchu, Taiwan

Wei-Jen Hsueh
Department of Electrical Engineering, National Central University, Chungli, Taiwan

Yi-Chen Chung
Department of Mechanical and Mechatronic Engineering, National Taiwan Ocean University, Keelung, Taiwan

Prof. Tzy-Rong Lin
Institute of Optoelectronic Sciences and Department of Mechanical and Mechatronic Engineering, National Taiwan Ocean University, Keelung, Taiwan (Plasmonic devices have advanced significantly in the past decade. Being one of the most intriguing devices, plamonic nanolasers plays an important role in biomedicine, chemical sensor, information technology, and optical integrated circuits. However, nanoscale plasmonic devices, particularly in ultraviolet regime, are extremely sensitive to metal and interface quality, which renders the development of ultraviolet plasmonics. Here, by addressing the material issues, we demonstrate a low threshold, high characteristic temperature metal-oxide-semiconductor ZnO nanolaser working at room temperature. The template for ZnO nanowires consists of a flat single-crystalline aluminum film grown by molecular beam epitaxy and an ultra-smooth $Al_2O_3$ spacer layer prepared by atomic layer deposition. By effectively reducing surface plasmon scattering loss and metal intrinsic absorption loss, the high-quality metal film and sharp interfaces between layers boost the device performance. Our work paves the way for future applications using ultraviolet plasmonic nanolasers and related devices.)

Plasmonic devices have caught much attention in the past decade due to its nanoscale size and high-speed operation. In recent years, various plasmonic devices have been proposed and demonstrated for many applications, such as gas detector [1], chemical sensor [2], photovoltaic devices [3, 4], biomedical sensor [5, 6], superlens [7, 8], optical trapping [9], optical tweezers [10], information technology [11, 12], optical integrated circuits [12-14] and nanoscale coherent emitter [15, 16, 17-20]. In these plasmonic devices, metal plays an important role for enhancing their performances. Gold, silver, and aluminum are frequently used metals to generate plasmons in visible and infrared spectral regime [3, 6, 8, 15, 18, 21-24]. In terms of device fabrication, gold is the best metal as it is stable in air. However, its interband transition around 2.3 eV makes it lossy and unsuitable for the applications aiming for wavelengths shorter than 540 nm [22, 25]. Silver has been applied for plasmonic devices in visible to ultraviolet regime. Due to its strong interband absorption for wavelengths less than ~350 nm, the intrinsic ohmic damping loss of silver is significant in ultraviolet regime. In comparison to gold and silver, simple-metal aluminum is not only a cost-effective and abundant metal material and widely used in modern semiconductor fabrication processes, but also has superior plasmonic properties than noble metals in the ultraviolet spectral regime. Furthermore, aluminum film is pretty stable in air and compatible with current CMOS technology, which is advantageous for integrating plasmonic devices with silicon-based electronic and photonic circuits [13, 26]. Therefore, it is widely accepted that aluminum is the most promising metal for ultraviolet plasmonics [17, 27]. On the other hand, it was demonstrated that ultra-smooth single-crystalline metal films are not only crucial for fabricating high-definition plasmonic nanostructures [22, 28], but also beneficial to the performance of fabricated optical antenna and plasmonic nanolasers [15, 18, 22].

Nevertheless, it has been a daunting challenge to grow larger-area, flat, and single-crystalline aluminum films as the growth is very sensitive to the surface condition and morphology. Molecular beam epitaxy (MBE) is one of the most promising methods to grow the wanted aluminum film because its ultra-high vacuum provides a very clean environment and the substrate native oxide can be desorbed in the chamber just before aluminum growth. Previously, several growth methods using MBE were carried out, such as on As-stabilized AlAs surface [29], thermal-anneal-induced [30] and grown on silicon (111) substrate [31]. However, the grown single-crystalline aluminum films, even up to 200-nm-thick, were pretty rough. The rough surface is an obstacle to development nanoscale plasmonic devices as it would increase surface plasmon scattering rate and reduce the propagation length. Herein, by using a minimum migration method [32], we have successfully grown a very flat single-crystalline aluminum films with root-mean-square roughness of 0.44 nm on GaAs substrate by MBE. The large-area growth conducted by Gallium-rich surface condition, flat GaAs surface morphology, and high growth rate is highly reproducible, which is an important step toward fabricating ultraviolet plasmonic devices.

We choose ultraviolet plasmonic nanolasers to demonstrate the crucial role of metal quality because they could provide a unique setting for the manipulation of light via the confinement of the electromagnetic field to a size well below the diffraction limit and they are an ideal nano-platform for investigating the light interacting with nano-materials [33, 34]. Furthermore, how to reduce plasmonic losses is a key issue for plasmonic nanolasers [17, 18, 22, 23], particularly in ultraviolet regime. The lasing wavelength is in nanoscale so the plasmonic losses are very sensitive to metal crystallization [18], lateral correlation length of grain boundaries [24] and surface morphologies [17]. Here, we first present that the MBE-grown aluminum film boost the nanolaser performance by minimizing metal intrinsic damping loss and scattering loss. Basically, the plasmonic lasing mechanism could be explained by surface plasmons amplification by stimulated emission radiation (SPASER) principle [35], analogous to the conventional photon lasers. The excitons in the gain medium are excited and nonradiatively transferred their energy to resonant surface plasmons SPs providing amplification channels. Due to the ultrasmall modal volume ($V_m$), typically small $V_m \sim \lambda^3/10 -$ $\lambda^3/1000$, of plasmonic cavity [15, 16, 17-19, 33, 36] is beneficial to obtain a high Purcell factor ($F_p$) to efficiently boost the energy transfer so the population inversion of excitons could be easily realized, which are highly advantageous for lowering its lasing threshold.

In this work, we use the metal-oxide-semiconductor (MOS) structure [15, 17-19] to demonstrate ultraviolet nanolasers on single-crystalline aluminum film. The key points to realize ultraviolet MOS nanolasers are listed in the following. First, a high-quality, single-crystalline metal film is crucial. It could efficiently reduce the metallic loss and provide high conductivity to enhance optical confinement and to prolong the SPs propagating length. Second, a flat surface morphology could drastically lower the SPs scattering loss. Finally, a close contact at planar metal/oxide and oxide/semiconductor interfaces greatly lessens the scattering loss, and more importantly, efficiently promotes the exciton-SP energy transfer thus furnishes adequate gain to compensate the loss and to achieve lasing [15, 17]. As we shall present, ultra-low threshold with high temperature stability room-temperature ultraviolet nanolasers have been obtained. Our plasmonic nanolaser is consisted with a zinc oxide (ZnO) nanowire lying on an aluminum film with an $Al_2O_3$ spacer layer to form the MOS structure. The extremely flat single-crystalline aluminum film, denoted as SC-Al hereafter, on a carefully treated GaAs surface was grown by molecular beam epitaxy system. For comparison, we also prepared a poly-crystalline Al film, denoted as PC-Al, deposited by e-gun evaporation (See supporting information for details). Furthermore, ultra-smooth $Al_2O_3$ spacer layer deposited on SC-Al and PC-Al films by atomic layer deposition (ALD) provides the crucial interface for achieving a flat and close contact between metal and semiconductor. The ZnO nanowires serve as the gain medium in ultraviolet regime as the large exciton binding energy and oscillator strength of ZnO are beneficial to the coupling between the excitons and SPs and to room-temperature operation [37]. Besides, the hexagonal cross-section of ZnO nanowires could provide a wanted close contact. Our experimental results reveal the importance of ultra-smooth, high-quality, and single-crystalline Al films and the flat and sharp interfaces between layers to realizing a low-threshold room-temperature ultraviolet nanolasers.

**Figures 1 (a) and (b)** show the 5 × 5 μm$^2$ bird-view atomic force microscope (AFM) images of the SC-Al and PC-Al films in air, respectively. The root-mean-square roughness of SC-Al is 0.44 nm, which is 5.2-fold smaller than 2.29 nm of PC-Al. It is clear that there are several spikes in the image of PC-Al film, which could induce serious scattering loss of fabricated nanolasers after insulator layer deposition. **Figure 1 (c)** shows the measured reflectivity spectra of the SC-Al and PC-Al films. In comparison to the PC-Al film, the SC-Al has the higher reflectivity in ultraviolet to near-infrared wavelength region. In particular, for the spectral window below 400 nm, the reflectance of the SC-Al film is obviously larger than that of the PC-Al one, probably due to the ultra-smooth surface morphology of SC-Al film reducing the absorption and random scattering caused by rough surface and grain boundaries [24]. **Figure 1 (d)** shows the high-resolution transmission electron microscope (HRTEM) image taken from the SC-Al film after the deposition of Al$_2$O$_3$ layer by ALD method. The upper layer is the Al$_2$O$_3$ spacer layer and the bottom one is the SC-Al film with a clear and sharp interface in between. The nearly perfect, atomic-like periodic array in the Al film reveals that the Al film is of highly quality and indeed single-crystalline. The inset to **Figure 1 (d)** shows the electron diffraction pattern in Al region. The clear hexagonal diffraction pattern without any observable side points indicates the faced-cubic-center (FCC) crystal structure of Al. In order to identify the crystal orientation of the Al films relative to the GaAs substrate, we performed X-ray diffraction (XRD) measurements on both Al films. **Figure 2 (a)** shows the low-incident-angle 2θ scanning XRD on the PC-Al film. As expected, multiple diffraction peaks at 38.5°, 44.7° and 78.2° respectively corresponding to Al (111), (200), and (311) surfaces are observed. On the other hand, we could not observe any Al peak from the SC-Al film. A longer thought would understand that it is not unusual to see no peak at all for a single-crystalline thin film because the lattice planes of Al may not be parallel to those of GaAs substrate. As reported previously [22], owing to the very close diffraction peaks between GaAs (100) and Al (110) surfaces, the later one could be buried in the strong signal of the former one. To avoid the interference from the GaAs substrate, we looked for the other lattice plane Al (111) that has no radial symmetry and carried out ϕ-dependent scanning using the experimental setup shown in **Figure 2 (b)**. Because the included angle between Al (111) and (100) plane is 34.5°, we set χ as 34.5° and 2θ as 38.5° for ϕ-scan. **Figure 2 (c)** shows the ϕ-dependent measurement result of the PC-Al film. As expected, no clear ϕ-dependence of counts is spotted due to its random crystal orientations. In contrast, **Figure 2 (d)** shows a clear and ϕ-dependent XRD peak of Al (111) of the SC-Al specimen. Note that the peak count of SC-Al is more than two-order larger than that of PC-Al, confirming that the SC-Al film is indeed single-crystalline and of high quality.

We expect that the performance of plasmonic devices in ultraviolet regime, such as ZnO-based MOS nanolasers, will be significantly enhanced by the superior quality of our SC-Al film. To demonstrate the surface roughness and metal crystallization are both crucial for plasmonic nanolasers, as the Figure 3 (a) shows, we put ZnO nanowires on both SC-Al and PC-Al films with a 5-nm-thick ultra-smooth $Al_2O_3$ dielectric spacer layer grown by ALD (supporting information). In comparison to PC-Al film, the SC-Al film could provide an ultra-smooth template for growing an ultra-flat $Al_2O_3$ spacer layer, which can provide proper optical confinement and prevent generated excitons in ZnO nanowires from fast quenching on the metal surface so the efficient exciton-SP energy transfer could sustain [7]. The scanning electron microscope (SEM) image of a finished ZnO nanolaser is shown in **Figure 3 (b)**. The ZnO nanowire length and the hexagonal side length are about 1 μm and 30 nm, respectively. **Figure 3 (c)** shows the HRTEM image of a ZnO nanowire taken from the $[2\bar{1}\bar{1}0]$ direction. The lattice fringe spacing in the HRTEM image is 0.52 nm, which corresponds to the lattice constant of the hexagonal wurtzite structure of ZnO and indicates that the nanowires grow along the [0001] direction.

**Figure 4** illustrates the optical measurements on two typical nanolasers on SC-Al and PC-Al films at 77 K. **Figure 4 (a)** shows the emission intensity and the linewidth versus pumping energy density for a 1-μm-long, single ZnO nanowire on $Al_2O_3$/SC-Al. It is clearly seen that, in the log-log plot of pumping energy density and emission intensity, a nonlinear S-shaped dependence appears and, in the transition region from ~0.2 – 0.4 mJ cm$^{-2}$, a dramatic reduction of emission linewidths (from ~ 8 to ~0.2 nm) occurs. This is a clear evidence of nanolaser lasing with an ultra-low threshold energy density of 0.28 mJ cm$^{-2}$. **Figure 4 (b)** shows the corresponding pumping energy density-dependent emission spectra. Below the lasing threshold, we can see the weak and broad emission spectra around 370 nm. When the pumping energy density increases to 0.38 mJ cm$^{-2}$, a very narrow lasing peak at 371 nm is obtained. For comparison, the nanolasers on PC-Al film was measured at 77 K and a typical result from a 1.46-μm-long ZnO nanowire is shown in **Figures 4 (c)** and **(d)**. Similarly, we can see a nonlinear S-shaped response behavior and linewidth narrowing down to 0.3 nm with a much larger threshold energy density of 10.19 mJ cm$^{-2}$, which is 36-fold larger than that on SC-Al. **Figure 4 (d)** shows the measured pumping energy density -dependent emission spectra. Below the lasing threshold, a broad spontaneous emission spectrum of the ZnO nanowire was obtained. Due to the high injection condition above threshold, that lasing peak at 372 nm is accompanied with a broad emission even at the pumping energy density of 12.73 mJ cm$^{-2}$. Our result reveals that the crystal quality of underneath metal film is indeed crucial for the device performance of the ZnO nanolasers as the metallic loss is minimized. Therefore, the SC-Al film could reduce the threshold condition to benefit the nanolaser operation.

To emphasize the importance of the metal and spacer layers, we further present the threshold and yield of more ZnO nanolasers on four kinds of templates: $Al_2O_3$ layer by ALD on PC-Al and on SC-Al films, and $SiO_2$ layer by e-gun evaporation on PC-Al and on SC-Al. The insulating spacer layers are all 5-nm thick and the preparation details can be found in the Supporting information. Characterized by AFM, the $SiO_2$ spacer layer deposited by e-gun evaporator has much rougher surface than the $Al_2O_3$ grown by ALD. Totally, about fifty nanolasers selected from the fabricated four chips were measured at 77 K to record the device yield and threshold pumping energy density. **Figure 5 (a)** shows numbers of measured (blue color bars) and lasing (red color bars) nanolasers. Not surprisingly, the best yield of 87% is obtained with the nanolasers on the SC-Al/$Al_2O_3$ template. The worst case is the devices on the PC-Al/$SiO_2$ template. Only one out of twelve devices is working. The second best template is SC-Al/$SiO_2$ indicating the key role of the metal film in plasmonic devices. **Figure 5 (b)** shows the corresponding threshold pumping energy density of all lasing devices at 77 K. The only one lasing nanolaser on the $SiO_2$/PC-Al templat has a very high threshold of 54.5 mJ/cm$^2$ (black circle). The threshold pumping densities of nanolasers on SC-Al/$SiO_2$ (red triangles) and PC-Al/$Al_2O_3$ (green diamond) templates are quite close to each other and in the range of 10 to 30 mJ cm$^{-2}$. The nanolasers on the SC-Al/ $Al_2O_3$ template have in general the lowest threshold pumping densities (blue star) ranging from ~0.28 to ~2 mJ cm$^{-2}$. Comparing with the nanolasers on other templates, the threshold pumping energy density is in general one to two orders lower. This experimental result unambiguously evidence the superior quality of our SC-Al/$Al_2O_3$ template prepared by MBE and ALD.

With the best template, our ZnO nanolasers can operate at even room temperature. **Figure 6 (a)** shows the measured lasing spectra of one nanolaser on SC-Al/ $Al_2O_3$ template from 77 – 300 K. With increasing temperature, the lasing behavior sustains up to 300 K and lasing wavelength red shifts from 371 to 381 nm. It is worth noting that the threshold pumping energy density increases from 0.28 mJ cm$^{-2}$ at 77 K to 0.84 mJ cm$^{-2}$ at 273 K corresponding to a quite large characteristic temperature ($T_0$) of about 178 K. The large $T_0$ could mainly arise from the high-quality template with low metallic losses. Due to the high Fermi temperature of metal, the nanolaser threshold increases with temperature probably due to the extra loss contributed from the electron-phonon scattering loss in metal and ZnO nanowires. **Figure 6 (b)** shows the emission intensity and linewidth versus pumping energy density of the 1-μm-long nanowire lying on SC-Al/Al$_2$O$_3$ template measured at 300 K. Clear lasing features, such as S-shaped response curve with threshold pumping energy density of 6.1mJ cm$^{-2}$–and dramatic linewidth reduction down to 0.3 nm, remain. **The inset of Figure 6 (b)** shows that the emission polarization is highly polarized along the direction parallel to the nanowire axis with a degree of polarization of 75%, which indicates that the longitudinal lasing modes is still the fundamental SP mode.

In conclusion, we have presented a high-performance plasmonic nanolaser in the ultraviolet regime. The interfacial roughness and, in particular, metal film quality play a key role in the ZnO nanolasers. By using molecular beam epitaxy to grow a high-quality single-crystalline Al film, followed by ultra-smooth Al$_2$O$_3$ layer prepared by atomic layer deposition and ZnO nanowire placement, we have realized an ultraviolet plasmonic nanolaser with a very low threshold pumping energy density and a high characteristic temperature. The nanolasers operated at room temperature shows clear features of lasing action. Our work reveals the importance of metal film quality and interface control for the plasmonic devices and paves the way for further applications using ultraviolet nanolasers.

## Methods:

**Sample preparation**

The ultraviolet plasmonic nanolasers were made by placing the ZnO nanowires on a 100 nm-thick Al film with a 5 nm-thick insulating spacer layer forming a metal-oxide-semiconductor (MOS) structure. For comparison, a single-crystalline Al (SC-Al) grown by MBE and a poly-crystalline Al (PC-Al) film deposited by e-gun evaporation were prepared as the initial templates. The growth details of both Al films can be found in the Supporting information. The Al$_2$O$_3$ and SiO$_2$ insulating spacer layers grown by ALD and e-gun evaporator, respectively.

**Optical experiment set-up**

For optical measurements, the nanolasers were placed in a cryogenic vacuum chamber, which can be cooled down from 300K to 77 K by liquid nitrogen. We used a charge-coupled device (CCD) camera to find those specific regions with single ZnO nanowires. An Nd:YVO$_4$ 355-nm pulsed laser with a pulse duration of 0.5-ns pulse duration and a repetition rate of 1 kHz serves as the pumping light source. The normally incident circular-polarized pumping beam was focused to a 15-μm diameter spot by means of a 100×, near-ultraviolet, infinity-corrected objective lens with a numerical aperture of 0.55. The emitted light from ZnO nanowires was collected by the same objective lens, coupled to a 600-μm core ultraviolet optical fiber, and then feed into a 320-mm single monochromator equipped with a liquid-nitrogen-cooled CCD to resolve the emission spectra with a spectral resolution of 0.2 nm. A rotating polarizer was placed in front of the fiber for measurement of the degree of polarization, defined by $(I_{max}-I_{min})/(I_{max}+I_{min})$ where $I_{max}$ and $I_{min}$ are the maximum and the minimum intensity of the lasing peak, respectively.

## Simulations:

The 2-D device simulations were performed by COMSOL RF module, we used the eigenvalue solver of the finite-element-method (FEM) to find eigenmodes of the ZnO nanowire lying on the dielectric/metal substrate. The computation domains were enclosed by perfectly matched layers (PMLs) to absorb the scattered power with minimum reflection. For simulation, the refractive index of ZnO, Al$_2$O$_3$ and SiO$_2$ are 2.54, 1.79 and 1.47, respectively. The simulation details and aluminum refractive index extracted from measured reflectivity spectra are provided in supporting information.

## Acknowledgements

Authors acknowledge the technical support from Profs. S. C. Wang and H. C. Kuo from National Chiao Tung University and the help from Mr. Chih-Kai Chiang at National Taiwan Ocean University for simulation support. This work was financially supported by the Ministry of Education Aim for the Top University program and by Minister of Science and Technology under Contract Nos. 104-2923-E-009 -003 -MY3, 102-2221-E-009 -156 -MY3 and 103-2221-E-019-028-MY3 in Taiwan.

[1]    B. Liedberg, C. Nylander, I. Lunström, Sensors and Actuators **1983**, 4, 299.
[2]    a) J. N. Anker, W. P. Hall, O. Lyandres, N. C. Shah, J. Zhao, R. P. Van Duyne, Nat Mater **2008**, 7, 442; b) M. E. Stewart, C. R. Anderton, L. B. Thompson, J. Maria, S. K. Gray, J. A. Rogers, R. G. Nuzzo, Chemical Reviews **2008**, 108, 494.
[3]    W.-L. Liu, F.-C. Lin, Y.-C. Yang, C.-H. Huang, S. Gwo, M. H. Huang, J.-S. Huang, Nanoscale **2013**, 5, 7953.
[4]    a) S. Mubeen, J. Lee, W.-r. Lee, N. Singh, G. D. Stucky, M. Moskovits, ACS Nano **2014**, 8, 6066; b) H. J. Park, L. J. Guo, Chinese Chemical Letters **2015**, 26, 419; c) H. A. Atwater, A. Polman, Nat Mater **2010**, 9, 205.
[5]    a) W. P. Hall, J. Modica, J. Anker, Y. Lin, M. Mrksich, R. P. Van Duyne, Nano Letters **2011**, 11, 1098; b) N. G. Khlebtsov, L. A. Dykman, Journal of Quantitative Spectroscopy & Radiative Transfer **2010**, 111, 1; c) D. O. Shin, J.-R. Jeong, T. H. Han, C. M. Koo, H.-J. Park, Y. T. Lim, S. O. Kim, Journal of Materials Chemistry **2010**, 20, 7241.
[6]    a) C. Huang, J. Ye, S. Wang, T. Stakenborg, L. Lagae, Appl Phys Lett **2012**, 100, 173114; b) M. Hu, J. Chen, Z.-Y. Li, L. Au, G. V. Hartland, X. Li, M. Marquez, Y. Xia, Chemical Society Reviews **2006**, 35, 1084.
[7]    a) Z. Liu, S. Durant, H. Lee, Y. Pikus, N. Fang, Y. Xiong, C. Sun, X. Zhang, Nano Letters **2007**, 7, 403; b) S. Kawata, Y. Inouye, P. Verma, Nat Photon **2009**, 3, 388; c) S. Huang, H. Wang, K.-H. Ding, L. Tsang, Opt. Lett. 2012, 37, 1295; X. Zhang, Z. Liu, Nat Mater **2008**, 7, 435.
[8]    N. Fang, H. Lee, C. Sun, X. Zhang, Science **2005**, 308, 534.
[9]    a) B. J. Roxworthy, K. D. Ko, A. Kumar, K. H. Fung, E. K. C. Chow, G. L. Liu, N. X. Fang, K. C. Toussaint, Nano Letters **2012**, 12, 796; b) C. Chen, M. L. Juan, Y. Li, G. Maes, G. Borghs, P. Van Dorpe, R. Quidant, Nano Lett **2012**, 12, 125.
[10]   a) A. N. Grigorenko, N. W. Roberts, M. R. Dickinson, ZhangY, Nat Photon **2008**, 2, 365; b) S. K. Mondal, S. S. Pal, P. Kapur, Optics Express **2012**, 20, 16180; c) M. L. Juan, M. Righini, R. Quidant, Nat Photon **2011**, 5, 349; d) M. Righini, G. Volpe, C. Girard, D. Petrov, R. Quidant, Physical review letters **2008**, 100, 186804.
[11]   P. Berini, I. De Leon, Nat Photonics **2012**, 6, 16.
[12]   M. T. Hill, M. Marell, E. S. Leong, B. Smalbrugge, Y. Zhu, M. Sun, P. J. van Veldhoven, E. J. Geluk, F. Karouta, Y.-S. Oei, Optics Express **2009**, 17, 11107.
[13]   B.-T. Chou, T.-C. Lu, S.-D. Lin, J. Lightwave Technol. **2015**, PP, 1. DOI:10.1109/jlt.2015.2397889
[14]   a) K. Ding, Z. C. Liu, L. J. Yin, H. Wang, R. B. Liu, M. T. Hill, M. J. H. Marell, P. J. van Veldhoven, R. Notzel, C. Z. Ning, Appl Phys Lett **2011**, 98, 231108; b) M. P. Nezhad, A. Simic, O. Bondarenko, B. Slutsky, A. Mizrahi, L. A. Feng, V. Lomakin, Y. Fainman, Nat Photonics **2010**, 4, 395.
[15]   Y.-H. Chou, B.-T. Chou, C.-K. Chiang, Y.-Y. Lai, C.-T. Yang, H. Li, T.-R. Lin, C.-C. Lin, H.-C. Kuo, S.-C. Wang, T.-C. Lu, ACS Nano **2015**, 9, 3978.
[16]   a) M. Noginov, G. Zhu, A. Belgrave, R. Bakker, V. Shalaev, E. Narimanov, S. Stout, E. Herz, T. Suteewong, U. Wiesner, Nature **2009**, 460, 1110; b) T. P. Sidiropoulos, R. Röder, S. Geburt, O. Hess, S. A. Maier, C. Ronning, R. F. Oulton, Nature Physics **2014**, 10, 870.
[17]   Q. Zhang, G. Li, X. Liu, F. Qian, Y. Li, T. C. Sum, C. M. Lieber, Q. Xiong, Nature communications **2014**, 5, 4953.
[18]   Y.-J. Lu, J. Kim, H.-Y. Chen, C. Wu, N. Dabidian, C. E. Sanders, C.-Y. Wang, M.-Y. Lu, B.-H. Li, X. Qiu, science **2012**, 337, 450.
[19]   R. F. Oulton, V. J. Sorger, T. Zentgraf, R.-M. Ma, C. Gladden, L. Dai, G. Bartal, X. Zhang, Nature **2009**, 461, 629.
[20]   D. Saxena, S. Mokkapati, P. Parkinson, N. Jiang, Q. Gao, H. H. Tan, C. Jagadish, Nat Photon **2013**, 7, 963.
[21]   R. Flynn, C. Kim, I. Vurgaftman, M. Kim, J. Meyer, A. Mäkinen, K. Bussmann, L. Cheng, F.-S. Choa, J. Long, Optics express **2011**, 19, 8954.
[22]   H.-W. Liu, F.-C. Lin, S.-W. Lin, J.-Y. Wu, B.-T. Chou, K.-J. Lai, S.-D. Lin, J.-S. Huang, ACS Nano **2015**.

[23] B.-T. Chou, S.-D. Lin, B.-H. Huang, T.-C. Lu, Journal of Vacuum Science and Technology B **2014**, 32, 031209.
[24] J. H. Park, P. Ambwani, M. Manno, N. C. Lindquist, P. Nagpal, S.-H. Oh, C. Leighton, D. J. Norris, Advanced Materials **2012**, 24, 3988.
[25] M. Fox, *Optical Properties of Solids*, Oxford, New York **2010**.
[26] a) A. D. Neira, G. A. Wurtz, P. Ginzburg, A. V. Zayats, Optics Express **2014**, 22, 10987; b) N. Pavarelli, J. S. Lee, M. Rensing, C. Scarcella, S. Zhou, P. Ossieur, P. A. O'Brien, J. Lightwave Technol. **2015**, 33, 991.
[27] J. M. McMahon, G. C. Schatz, S. K. Gray, Physical Chemistry Chemical Physics **2013**, 15, 5415.
[28] J.-S. Huang, V. Callegari, P. Geisler, C. Brüning, J. Kern, J. C. Prangsma, X. Wu, T. Feichtner, J. Ziegler, P. Weinmann, M. Kamp, A. Forchel, P. Biagioni, U. Sennhauser, B. Hecht, Nat Commun **2010**, 1, 150.
[29] N. Maeda, M. Kawashima, Y. Horikoshi, Journal of Applied Physics **1993**, 74, 4461.
[30] H. F. Liu, S. J. Chua, N. Xiang, Journal of Applied Physics **2007**, 101, 053510.
[31] N. Joshi, A. K. Debnath, D. K. Aswal, K. P. Muthe, M. Senthil Kumar, S. K. Gupta, J. V. Yakhmi, Vacuum **2005**, 79, 178.
[32] a) A. Y. Cho, P. D. Dernier, Journal of Applied Physics **1978**, 49, 3328; b) S.-W. Lin, J.-Y. Wu, S.-D. Lin, M.-C. Lo, M.-H. Lin, C.-T. Liang, Japanese Journal of Applied Physics **2013**, 52, 045801.
[33] D. K. Gramotnev, S. I. Bozhevolnyi, Nat Photon **2010**, 4, 83.
[34] a) M. S. Tame, K. R. McEnery, S. K. Ozdemir, J. Lee, S. A. Maier, M. S. Kim, Nat Phys **2013**, 9, 329; b) J. Takahara, S. Yamagishi, H. Taki, A. Morimoto, T. Kobayashi, Opt. Lett. **1997**, 22, 475.
[35] D. J. Bergman, M. I. Stockman, Physical review letters **2003**, 90, 027402.
[36] R.-M. Ma, R. F. Oulton, V. J. Sorger, G. Bartal, X. Zhang, Nature materials **2011**, 10, 110.
[37] Y.-Y. Lai, Y.-P. Lan, T.-C. Lu, Light Sci Appl **2013**, 2, e76.

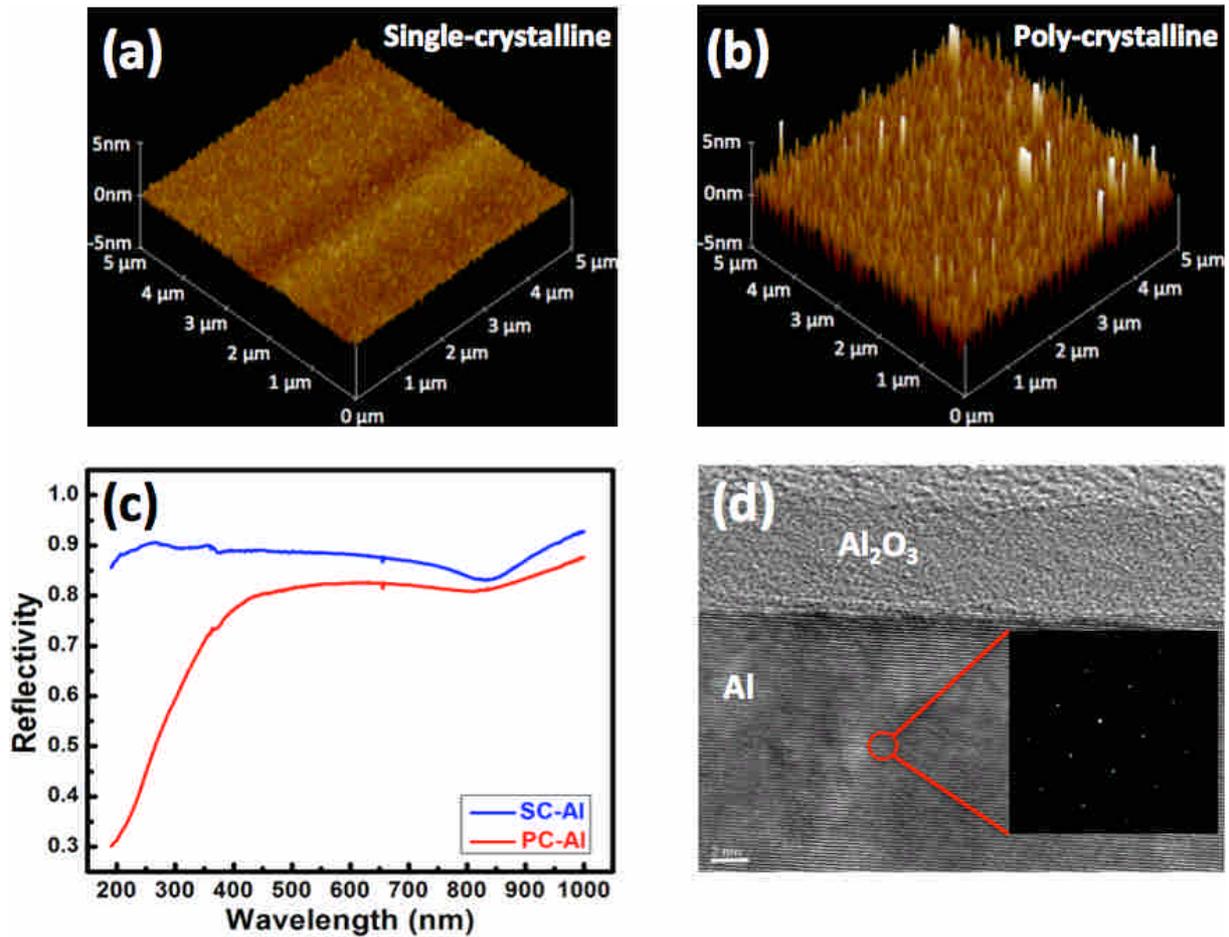

**Figure 1.** 5 × 5 μm$^2$ bird-view atomic force microscope images of the single- and poly-crystalline Al films in (a) and (b), respectively. (c) Reflectivity spectra of both Al films. (d) Cross-sectional transmission electron microscopy images of Al$_2$O$_3$ layer on single-crystalline Al film. Inset in (d): electron diffraction pattern from the Al region.

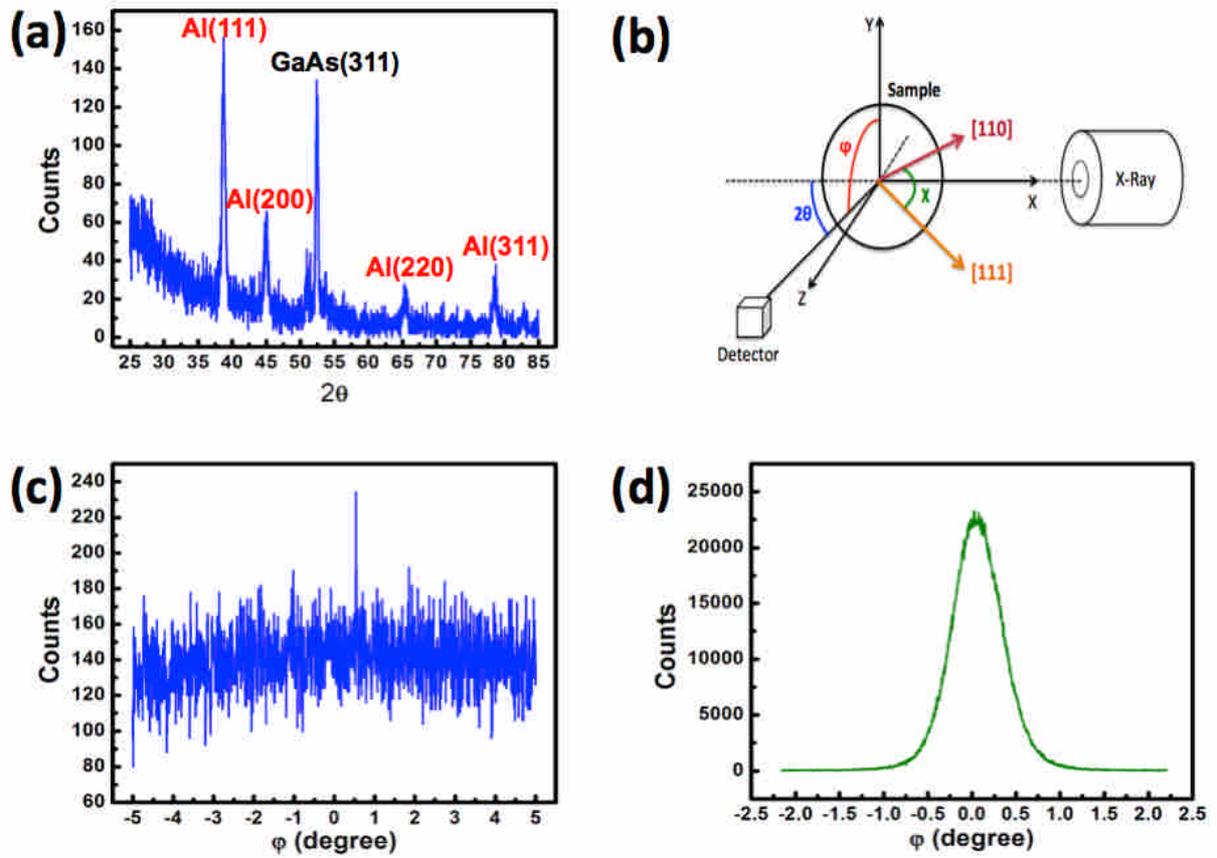

**Figure 2.** (a) Low-incident-angle 2θ-scanning of poly-crystalline Al film. (b) XRD setup for φ-dependent measurement of Al-(111) plane (c) Measured φ-dependent scanning of poly-crystalline Al film. (d) Measured φ-dependent scanning of single-crystalline Al film.

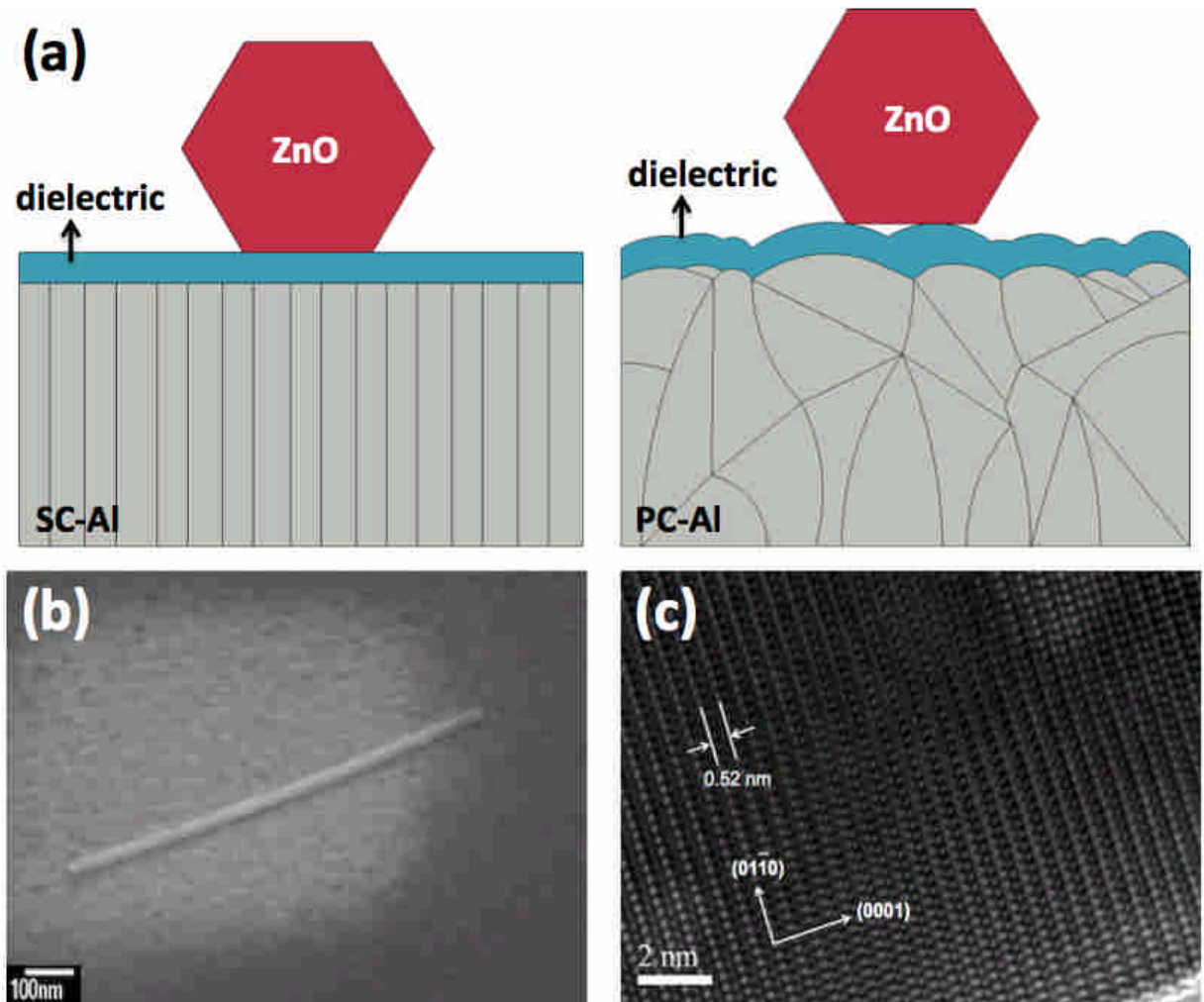

**Figure 3.** (a) Schematics of a ZnO nanowire lying on the top of single- and poly-crystalline Al films with a dielectric spacer layer. (b) Scanning electron microscope image of a ZnO plasmonic nanolaser. The ZnO nanowire length and the hexagonal side length are 1 μm and 30 nm, respectively. (c) High-resolution transmission electron microscopy image of a ZnO nanowire taken from $[2\bar{1}\bar{1}0]$ direction.

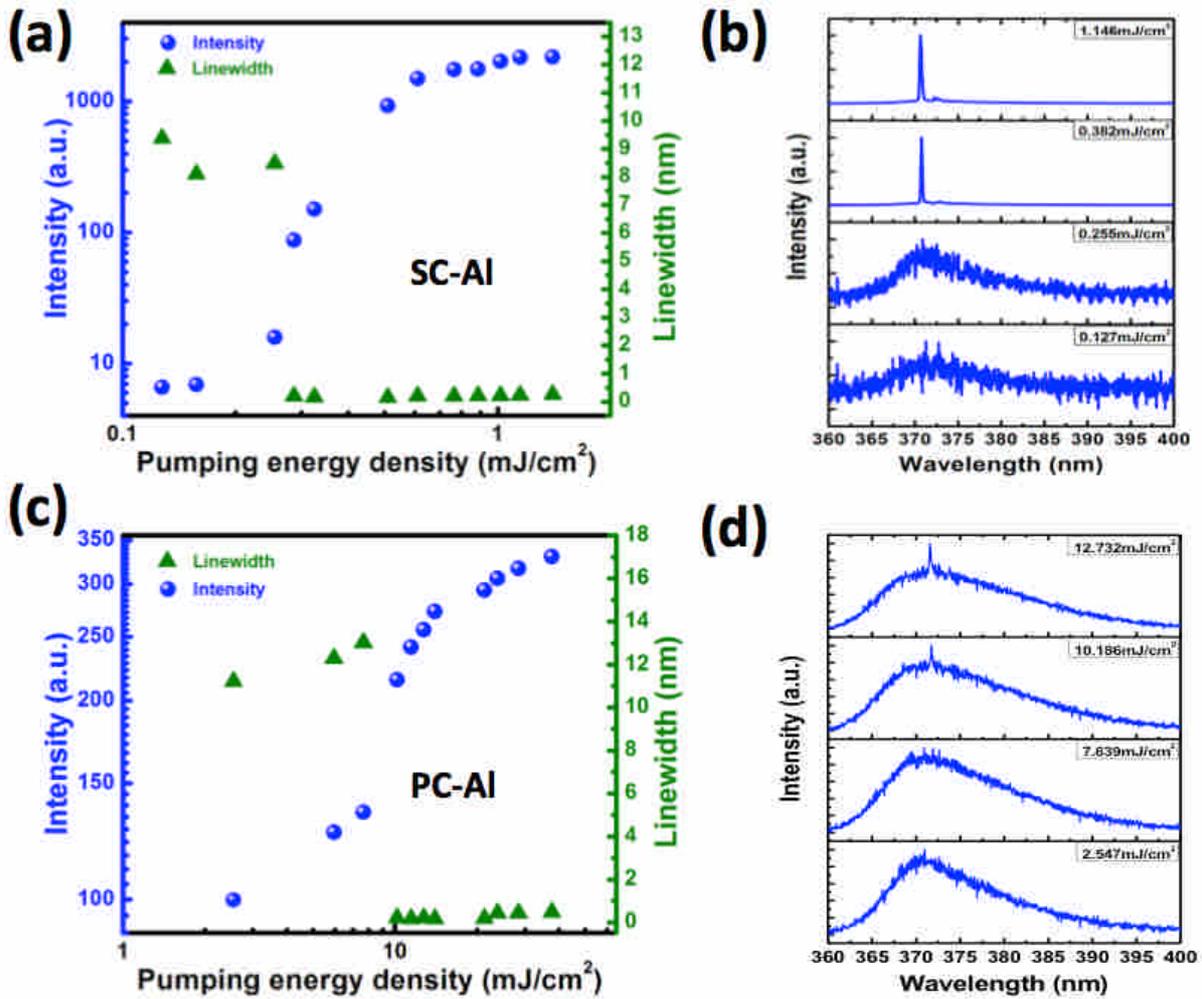

**Figure. 4.** Emission intensity and linewidth as a function of pumping energy density in (a) for a nanolaser on SC-Al/Al$_2$O$_3$ template and the corresponding emission spectra in (b). Emission intensity and linewidth as a function of pumping energy density in (c) for a nanolaser on PC-Al/Al$_2$O$_3$ template and the corresponding emission spectra in (d). All data were measured at 77 K.

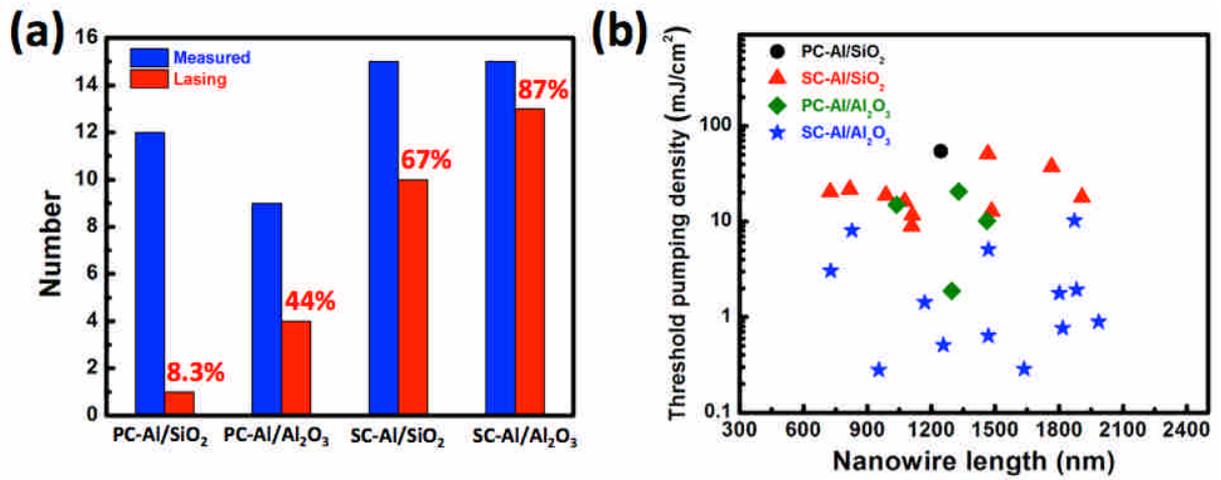

**Figure. 5.** (a) Numbers of measured and lasing nanowires for the four kinds of specimens. (b) Threshold pumping energy density as a function of nanowire length for the four kinds of specimens. All data were measured at 77 K.

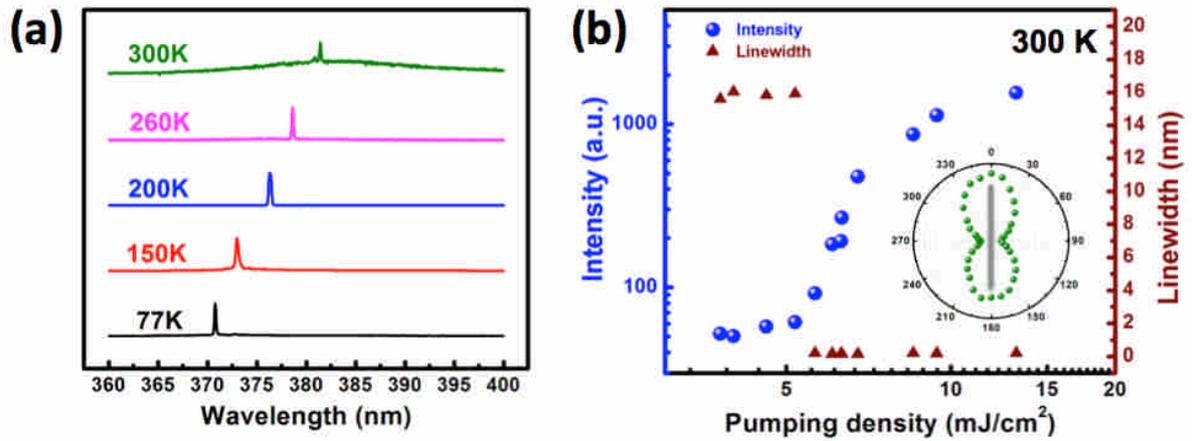

**Figure. 6.** (a) Temperature-dependent lasing spectra of a nanolaser on SC-Al/Al$_2$O$_3$ template from 77 K to 300 K. (b) Emission intensity and linewidth versus pumping energy density of the nanolaser at 300 K. Inset on (b): corresponding lasing polarization plot.